\documentclass[12pt]{article}

\usepackage{graphics}
\input{amssym.def}
\input{amssym.tex}
\normalbaselines
\begin{document}
\begin{center}

\vspace*{1.0cm}

{\LARGE{\bf Aspects  of Gauge Theory - Gravity\\[2mm] Correspondence}}

\vskip 1.5cm

{\large {\bf S. Randjbar-Daemi}}

\vskip 0.5 cm

the Abdus Salam International Centre for Theoretical Physics \\
Strada Costiera 11\\
Trieste, Italy
\end{center}

\renewcommand{\thefootnote}{\fnsymbol{footnote}}

\vspace{1cm}

\begin{abstract}
 A brief review of aspects of gravity gauge theory correspondance inspired
 by string theory is presented. \footnote{Based on invited talks given at the International
Symposium on Quantum Theory and Symmetries (8-22 July 1999, Goslar, Germany) and
the Sixth International Wigner Symposium (WIGSYM6) (16-22 August 1999,
Istanbul, Turkey).}

\end{abstract}

\vspace{1 cm}

\renewcommand{\thefootnote}{\arabic{footnote}}
\setcounter{footnote}{0}

\noindent{\bf 1. Introduction}

Recent developments in superstring theory have opened up new avenues for closer
 connections of this theory with 4-dimensional physics. Of particular
significance is our better
 understanding of the physics of  extended objects, or branes,  which
appear as classical
solutions in  string theories.
 Studies  of  $D$-brane physics in particular  have given us the new
perspective
of regarding  open strings as the excitation states of these
configurations [1].
This new insight requires a very deep and still not well understood
interplay between
 Yang--Mills and gravitational interactions.

 One of the main ideas underlying  recent excitements is a conjecture  of
Maldacena [2], known as the $AdS/CFT$ correspondance,
which suggests a new connection between  gravitational and Yang--Mills forces.
This conjecture states that the strong coupling limits of certain gauge
theories, a
limit in which
the conventional perturbative methods fail, are dual to type IIB string theory
on a background geometry of $AdS_5\times X_5$, where $X_5$ is an Einstein
manifold
with a cosmological constant $\Lambda =-\Lambda_{AdS_5}$.  The case of $X_5
=S^5$
has been studied extensively. In this case the dual gauge theory is the
${\cal N}=4$ super Yang-Mills theory in $D=4$ with the gauge group $U(N)$.
Note that the strong coupling here refers to the large values of
the 'tHooft coupling $\lambda=g^2_{{\rm YM}}N$.
We are thus dealing with large $N$ gauge theory in which the leading
order term corresponds to planner Feynman graphs with spherical topology.
The subleading terms correspond to surfaces of higher genus [3].

Stringy excitations in gauge theories have often been assumed to be
responsible for
confinement of quarks in QCD. This picture becomes more plausible in the
lattice formulation
of Yang-Mills theory.  The existence of stringy
solutions in gauge theory  supports  this idea, too.
Another, perhaps stronger, evidence for a stringy regime in gauge
theories is obtained
in the large $N$ expansion . In this approach the gauge theory
Feynman graphs go
over  to string
loop graphs, i.e. in a sense the Feynman perturbation theory of a gauge
theory with
infinite number of colours gives rise to string perturbation theory.
The two dimensional Yang-Mills partition function is known
 to be equivalent
to a quasi topological closed string theory in the limit of large
$N$ [4].  On the other
hand, it is also well known that  all consistent string theories do  contain
gravity in their perturbative spectrum. It becomes plausible then to
speculate that there
might be  a deep interconnection between
gauge theories and
the gravitational physics. Infact recent developments give a reason why all
the
previous attempts at deriving a convincing string picture from a gauge theory
lagrangian have failed. One thing which the $AdS/CFT$ correspondance have
 taught us
is that the strong coupling desctription
of gauge theory requires infinite number
of new degrees of freedom provided by the
gravitational modes of the dual theory.

In this contribution, after briefly reviewing some elementary
facts about large $N$ gauge theory and the geometry of
conifolds,  we give a short description of the original $AdS/CFT$
correspondence
with the maximal supersymmetry. We shall then elaborate on some new
developments with
lesser
amount of supersymmetries. More specifically, we shall discuss $D_3$-brane
configurations
for which the transverse space is a non-compact Calabi-Yau  3-fold which is
known as a conifold. Klebanov and Witten have shown that this configuration
leads to a ${\cal N}=1$ superconformal theory with the gauge group
$SU(N)\times SU(N)$, where $N$ is the number of  $D_3$-branes [5]. In a sense
this theory
offers a better test of the Maldacena conjecture than the example of the
$AdS_5\times S_5$
background. In this latter case  the maximal symmetry of the background
 dictates the spectrum and  the structure of some of the Green functions.
The conifold background, on the other hand, has the minimum amount of conformal
 supersymmetries and therefore
the matching of  Yang-Mills states on the boundary and the Kaluza-Klein
states in the
bulk is not entirely dictated  by group theory.

 Conifolds are ubiquitous in string theory. They appear in the  Calabi-Yau
 compactifications of string theory [6] and the  effective action governing
low energy physics of moduli fields [7].
 We shall elaborate briefly on these points and some
others in section 3.

 Conifolds also enter the $c=1$ non critical string theory [9]. The partition
function of this theory happens to be equal to the coefficient of
certain terms in the
the Calabi-Yau compactifications of type IIB superstrings near a
conifold singularity [8]. One can regard the type IIB string near
a conifold point as the infinite $r$ limit of
a $D_3$ brane configuration. Starting from this infinte $r$ configuration
we can continousely approach the throat region near $r=0$, where
the $D_3$ brane geomtery factorises into $AdS_5\times T^{11}$.
\footnote{
The geomtery of $T^{11}$ will be explained in section 5.}
It is in the
background of this throat region that the Klebanov -Witten $\cal {N} =1$
superconformal
gauge theory is dual to the type IIB supergravity.
One may then ask
 about a
possible role
for the $c=1$ theory in the boundary Yang-Mills theory which arises in the
$AdS/CFT$
correspondace. This point will be touched upon in section [6].
\bigskip\bigskip

{\bf 2.\quad Conical Singularity} [6]

\bigskip

Consider a Calabi-Yau three folds defined by an algebraic equation

$$F(\xi_1,\dots,\xi_4) = 0$$

A nodal singularity is defined to be a point on the surface at which the
first derivatives
of $F$ vanish, viz;

$${\partial F\over \partial \xi_i} = 0\quad\qquad i = 1,\dots,4$$

Near a nodal point one can approximate the above algebraic equation by a
quadratic

$$\xi^2_1 + \dots +\xi^2_4 = 0\quad\qquad \xi_i\in \hbox{\Bbb C}$$

The surface defined by this equation will be denoted by  $Y_6$.
Note  that if $\vec\xi \in Y_6 $ then $ \lambda \vec\xi \in Y_6$
 for any $\lambda\in {\Bbb C}$.
 Thus $Y_6$ is the space of lines through the origin. We thus have a cone with
its  apex at $\xi_1 = \dots =\xi_4 = 0$.

  It is convenient to introduce real coordinates
$\vec \xi = \vec x+ i\vec y$ where $ \vec x\ \&\ \vec y \in {\Bbb R}^4$.
In these coordinates the equation of surface becomes
$$\vec x^{\phantom{.}2}-\vec y^{\phantom{.}2} = 0$$
$$\vec x\cdot \vec y = 0$$

The  base of the cone is obtained by intersecting the surface
 $Y_6$ with an $S^7$ centered at the apex
$$\vec x^{\phantom{.}2}+\vec y^{\phantom{.}2}=r^2$$
 We thus obtain for the base
$$\left\{
\begin{array}{l}
\vec x^{\phantom{.}2} =\vec y^{\phantom{.}2} ={r^2\over 2}\\[5mm]
\vec x\cdot\vec y = 0
\end{array}
\right.$$
Topologically this is $S^3\times S^2$,
\newpage

\setlength{\unitlength}{1mm}
\begin{center}
\begin{picture}(50,50)(60,-40)
\put(87,-2){Apex}
\put(81,-4.5){$\bullet$}
\put(82,-3){\line(1,-2){16}}
\put(82,-3){\line(-1,-2){16}}
\put(82,-3){\line(2,-1){25.5}}
\put(66,-35){\line(1,0){32}}
\put(107.5,-16){\line(-1,-2){9.5}}
\put(105,-30){$S^2$}
\put(80,-40){$S^3$}
\multiput(82,-16)(0,5){3}{\line(0,1){3}}
\multiput(82,-16)(5,0){5}{\line(1,0){3}}
\put(66,-35){\rotatebox{50}{\multiput(0,0)(5,0){5}{\line(1,0){3}}}}
\end{picture}
\end{center}


 The singularity at the apex can be resolved in two different ways
by replacing the singular point either with a $S^2$ (small resolution)
 or  with a $S^3$ (deformation).
Symbollically these two possibilities can be represented as in the
following figures:

\setlength{\unitlength}{0.7mm}

\begin{center}
\begin{picture}(50,50)(80,-30)
\put(25,-5){$S^2$}
\put(65,-40){$S^2$}
\put(35,-53){$S^3$}
\put(25,-45){\line(0,1){30}}
\put(25,-45){\line(1,0){30}}
\put(25,-15){\line(1,-1){30}}
\put(55,-45){\line(1,1){20}}
\put(25,-15){\line(1,1){20}}
\put(45,5){\line(1,-1){30}}
\put(25,-68){Small resolution}
\multiput(45,-25)(0,5){6}{\line(0,1){3}}
\multiput(45,-25)(5,0){6}{\line(1,0){3}}
\put(25,-45){\rotatebox{45}{\multiput(0,0)(5,0){6}{\line(1,0){3}}}}
\end{picture}

\begin{picture}(50,50)(-20,-70)
\put(65,-40){$S^2$}
\put(35,-52){$S^3$}
\put(53,17){$S^3$}
\put(25,-45){\line(1,0){30}}
\put(55,-45){\line(1,1){20}}
\put(75,-25){\line(0,1){40}}
\put(45,15){\line(1,0){30}}
\put(25,-45){\line(1,3){20}}
\put(55,-45){\line(1,3){20}}
\put(30,-68){Deformation}
\put(30,-80){$\xi^2_1+\dots + \xi^2_4 = \mu^2$}
\multiput(45,-25)(0,5){8}{\line(0,1){3}}
\multiput(45,-25)(5,0){6}{\line(1,0){3}}
\put(25,-45){\rotatebox{45}{\multiput(0,0)(5,0){6}{\line(1,0){3}}}}
\end{picture}
\end{center}

{\bf 3. Conifolds in String Theory}

 In  standard perturbation theory one expands each amplitude in a power
series
of a running coupling constant. If we assume that the number of colours,
$N$,  is
large the parameter  $1/N$ becomes small and therefore can be used as an
expansion parameter.
'tHooft [3] has shown that in this expansion the Yang-Mills Feynman
graphs look very
much like the string theory diagrams. To see this, it is convenient to adopt
a double line
notation in which any object in the adjoint representation of the gauge
group $U(N)$
is represented by two lines with arrows on them. Thus the gauge field $A$
is denoted by
$A^a_{\,b}$ $\begin{picture}(0,0)
\put(0,3){\line(1,0){10}}
\put(0,1){\line(1,0){10}}
\put(3,2){\hbox{\small $<$}}
\put(4,0){\hbox{\small $>$}}
\end{picture}$

 Using this notation the ordinary Feynman graphs like

$$\includegraphics{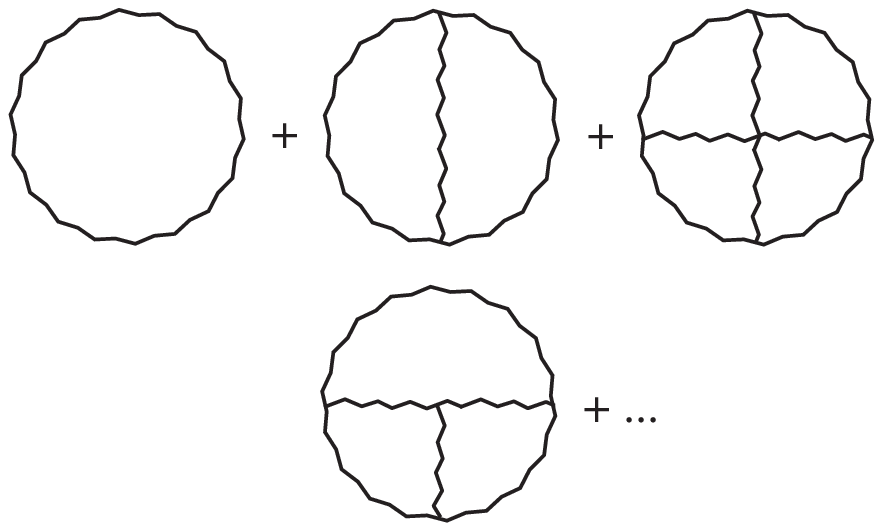}$$

\noindent
will look like  triangulations of a 2-dimensional surface,

$$\includegraphics{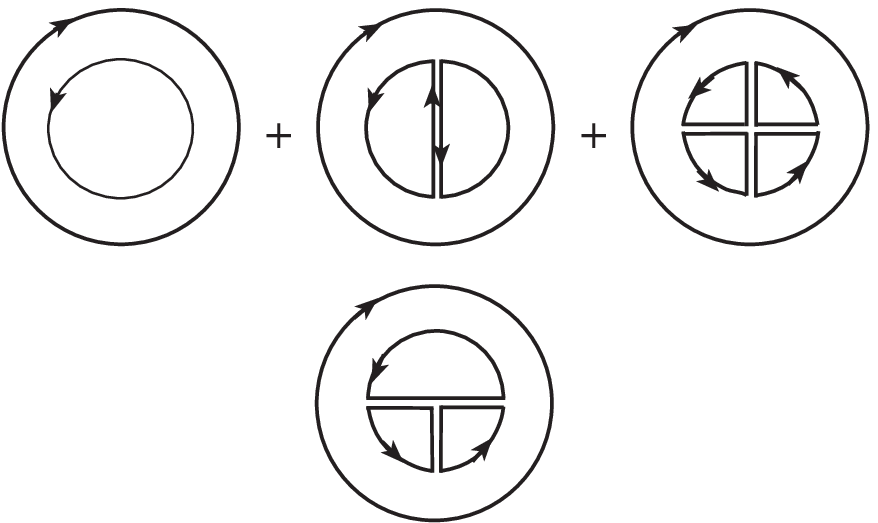}$$

This mapping  can be made more precise. To this end consider the Yang-Mills
lagrangian
$${\cal L} ={1\over g^2_{_{\rm YM}}}\left[-{1\over 4} {\rm Tr}
F^2+\dots\right] \ .$$

 To apply the ${1\over N}$ expansion  we need to keep the 'tHooft coupling
constant
$\lambda = g^2_{_{\rm YM}} N$  fixed as $N\mapsto \infty$. With this
redefinition
${\cal L}$ looks like
\bigskip\bigskip
$${\cal L} ={N\over \lambda} \left[-{1\over 4} {\rm Tr} F^2 +\dots\right]
\ .$$

 We can now read the $N$- and $\lambda$-dependence of each graph.
We note the following
sources of $N$-dependence in each graph: a factor of ${\lambda\over N}$
coming from each propagator,
 a factor of ${N\over\lambda}$ coming from each vertex,  and a factor
of $N$ coming from each loop. Therefore, a general graph with $h$ loops
(faces), $v$ vertices and  $p$
propagators will have the following $N$-dependence

$$ N^h \left({\lambda\over N}\right)^p\left({N\over \lambda}\right)^v =
N^{h-p+v}\lambda^{p-v} \ . $$

 The power of $N$ is the Euler number of a two dimensional surface
which is defined by

$$\chi = h-p+v$$
$$=2-2g$$
 where $g$ is the  topological genus of the surface.  If we identify  $1/N$
with a string theory
coupling
constant the Yang-Mills
perturbative expansion will go over to a topological expansion in powers of
the genus of
two dimensional surfaces which is characteristic for perturbative expansion in
 string theory. Note that the
partition
function of
the Yang-Mills theory, which is obtained from the summation of all the
Yang-Mills vacuum graphs,
will have the general form of
$$ \sum_{g=0\atop h=1}  C_{g,h} N^{2-2g} \lambda^{2g-2+h}$$
where $ C_{g,h}$ indicate the result of loop integration and other
algebraic operations on
each Feynman diagram in the Yang-Mills theory.
This expression can be given both an open string as well as a closed string
interpretations.

\bigskip
\noindent
1) {\bf Open string  interpretation}
\bigskip

The above result for the gauge theory  partition function indicates that
with each gauge theory diagram with $h$ loops, $v$ vertices and $p$ propagators
is associated a factor of $N^h (g{^2}_{YM})^{p-v}$. With an
appropriate interpretation of the parameters this factor becomes identical
to the one which one associates to an open string diagram  on a world sheet
with $g$ handles and $h$ boundaries. \footnote { For the Chern-Simons theory
we should replace $g^2_{YM}$ with ${ {2\pi N}\over {N+k}}$, where $k$ is the
 Chern-Simons coupling constant.}
 At a more intuitive level it is seen that
 the open string diagrams with  $g$ handles and $h$ boundaries like

$$\includegraphics{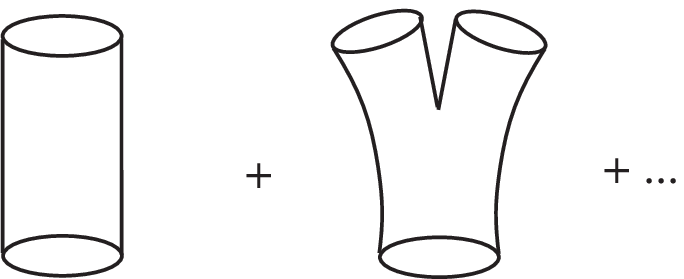}$$

\noindent
 can be mapped to the gauge theory Feynman graphs simply by flattening them,

$$\includegraphics{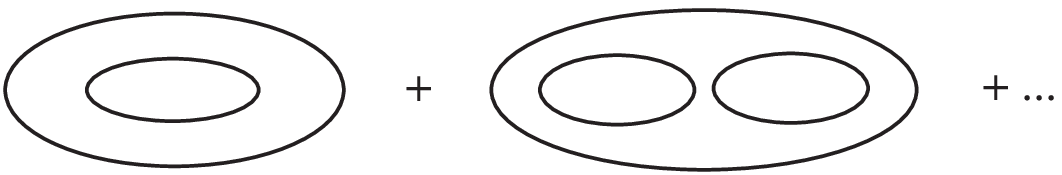}$$

A concrete realization of this
qualitative picture has been given by Witten [10] . Witten considers a
gauge
theory of Chern-Simons
type on $S^3$ with the partition function

$$Z={\displaystyle\int} [DA] e^{k\over 4\pi}
\mathop{\displaystyle\int}\limits_{S^3} {\rm Tr} (AdA+A^3)$$
and shows that the $C_{g,h}$ corresponding to this partition
function equals the partition function of a topological
open string theory on a world sheet with $g$ handles and
$h$ boundaries. Moreover the target space of this string theory, which is
nothing but the topological $A$ model,
is the cotangent bundle $T^*S^3$ of $S^3$    in a background
of $N$ $D_3$ branes wrapped on a $S^3$ submanifold.
The
cotangent
bundle of $S^3$ is  isomorphic to the manifold of the group $SL(2,C)$ which
is defined by

$$T^*S^3:XY-UV=1$$

 This manifold in turn is isomorphic  to

$$XY-UV=\mu^2$$

which is a deformed conifold.

\bigskip
\noindent
2) {\bf Closed string interpretation}
\bigskip

 Assume that we can perform the sum over $h$, i.e. number of holes on
the open string world sheet. The partition function then reduces to
a sum over $g$ of the form
$${\displaystyle\sum}_{g=0} N^{2-2g} {\rm F}_g(\lambda) \ .\eqno(1)$$

 This can be interpreted as a closed string perturbative expansion with the
string coupling
constant $g$ and the string (length) $^2$  $\alpha'$ identified as
$${1\over N}\sim g_s\ ,\quad \lambda \sim {1\over \alpha'} \ .$$

 Note that $N\to\infty$ implies $g_s\to 0$, i.e. string weak coupling
limit. This is the
perturbative regime in string theory. Thus the large $ N$ limit
of Chern-Simons theory maps to string perturbative
regime, as argued above.
It also becomes plausible to assume  that summing over the number of
boundaries of the
open string theory produces closed surfaces of the closed string theory.

A concrete realization of such resummation has been performed recently by
Gopakumar and Vafa [11]. These authors show that the partition function
of the
Chern-Simons theory on $S^3$ can indeed be mapped to a topological string
propagating
 on a small resolution of the conifold.


Graphically it seems that we have the following type of transition
$$\includegraphics{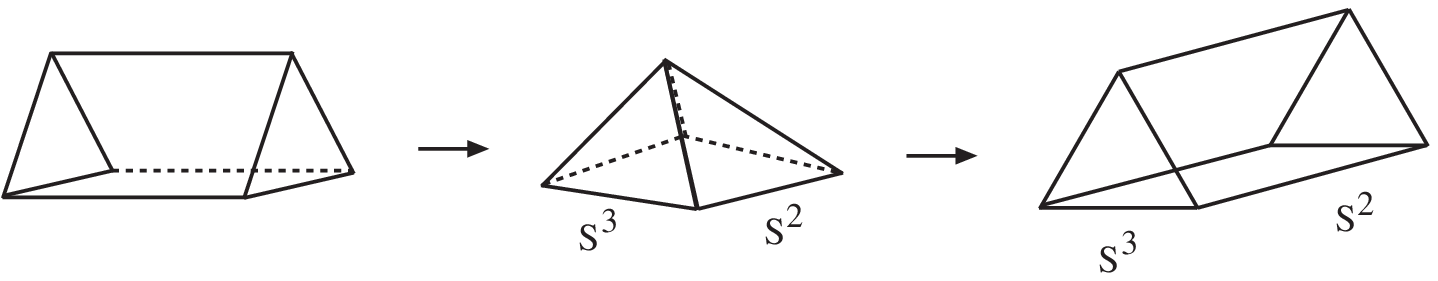}$$

\vspace{-2cm}

{\normalsize
$$\begin{tabular}{lllll}
Open strings on && The singular geometry&\phantom{000}& Closed strings\\
$T^*S^3 w/3$ branes && of  the conifold &&on $S^2$ resolved geometry\\
on $S^3\sim$ Chern-Simons &\phantom{000}&&&$(\lambda,N)\to (i\lambda,g_s
=\lambda/N)$\\
gauge theory
\end{tabular}$$}

  The Gopakumar-Vafa mapping starts from a detailed analysis of the
Chern-Simons partition function on $S^3$ which is known to be

$$\begin{array}{rcl}Z(S^3, N,h)&\equiv& e^{-F(S^3,N,k)}=\\[3mm]
&=&{1\over (N+k)^{N/2}} \prod^{N-1}_{j=1}
\left\{ 2\sin \left({j\pi\over N+k}\right)\right\}^{N-j}\ .\end{array} $$

Expanding this expression in powers of $1/N$ produces
$$F=\sum_{g=0} N^{2-2g} F_g(\lambda)$$
where $\lambda$ is defined by
$$
\lambda ={2\pi N\over N+ k} \ .$$
 One can  now use  the explicit form of $F(S^3,N,k)$ to
 calculate $F_g(\lambda)$. For example the $g=0$ term is given by

$$N^2F_0(\lambda)
= -\left({N^2\over\lambda}\right)\left[-\zeta (3) +{i\pi^2\over 6} \lambda
-\left(m+{1\over 4}\right)\pi\lambda^2 +{i\lambda^3\over 12} +
\sum^\infty_{n=1}
{e^{-in\lambda}\over n^3}\right] \ .
$$

With the substitution of  $g_s={i\lambda\over N}$ and
$ t = i\lambda$ this goes over to
$${1\over g^2_s}
\left[ -\zeta (3) +{\pi^2\over 6} t + i\left(m+{1\over 4} \right)\pi t^2
-{t^3\over 12} +\sum^\infty_{n=1} {e^{-nt}\over n^3}\right] \ .$$

This result agrees with $F_0(g_s,t)$ for a closed string on a $S^2$ resolved
conifold.
The $\sum^\infty_{n=1} {e^{nt}\over n^3}$ terms come from the world sheet
instantons.
Similar match have been obtained by Gopakumar and Vafa for $g\geq 1$
partition functions as well.
At least in the context of this simple setting it seems that a gauge theory
of Chern-Simons type can be
considered as a non perturbative version of gravity.

The closed and open string interpretation of the same Chern-Simons theory
is rather similar to the duality in AdS/CFT correspondence which we
 are going to briefly discuss in the
next paragraph. As  Gopakumar and Vafa suggest one can think of  $S^2$
as the two sphere surrounding the positions of the $D_3$ branes
in a transverse $R^3$ subspace inside $T^*S^3$.

Historically conifold entered string theory in connection with the Calabi-Yau
compactification from ten to four dimendions. For example
it has been observed by Candelas and de la Ossa that, in the context
of string theory, it is possible to make transitions between Calabi-Yau
threefolds of different topologies by passing through conifold points in
 the moduli space.  Graphically one can represent this process by
[6].

\setlength{\unitlength}{0.7mm}

\begin{center}
\begin{picture}(50,50)(100,-30)
\put(20,-5){$S^2$}
\put(70,-40){ $S^2$}
\put(35,-58){ $S^3$}
\put(25,-45){\line(0,1){30}}
\put(25,-45){\line(1,0){30}}
\put(25,-15){\line(1,-1){30}}
\put(55,-45){\line(1,1){20}}
\put(25,-15){\line(1,1){20}}
\put(45,5){\line(1,-1){30}}
\multiput(45,-25)(0,5){6}{\line(0,1){3}}
\multiput(45,-25)(5,0){6}{\line(1,0){3}}
\put(25,-45){\rotatebox{45}{\multiput(0,0)(5,0){6}{\line(1,0){3}}}}
\put(25,-70){ $\chi+2N$}
\end{picture}

\begin{picture}(50,50)(60,-70)
\put(87,0){Apex}
\put(80.5,-5){\hbox{\normalsize{$\bullet$}}}
\put(82,-3){\line(1,-2){16}}
\put(82,-3){\line(-1,-2){16}}
\put(82,-3){\line(2,-1){25.5}}
\put(66,-35){\line(1,0){32}}
\put(107.5,-16){\line(-1,-2){9.5}}
\put(105,-35){$S^2$}
\put(80,-50){$S^3$}
\put(74,-60){$\chi+N$}
\multiput(82,-16)(0,5){3}{\line(0,1){3}}
\multiput(82,-16)(5,0){5}{\line(1,0){3}}
\put(66,-35){\rotatebox{50}{\multiput(0,0)(5,0){5}{\line(1,0){3}}}}
\put(180,-40){$S^2$}
\put(145,-55){$S^3$}
\put(163,17){$S^3$}
\put(135,-45){\line(1,0){30}}
\put(165,-45){\line(1,1){20}}
\put(185,-25){\line(0,1){40}}
\put(155,15){\line(1,0){30}}
\put(135,-45){\line(1,3){20}}
\put(165,-45){\line(1,3){20}}
\multiput(155,-25)(0,5){8}{\line(0,1){3}}
\multiput(155,-25)(5,0){6}{\line(1,0){3}}
\put(135,-45){\rotatebox{45}{\multiput(0,0)(5,0){6}{\line(1,0){3}}}}
\put(143,-65){$\chi$}
\end{picture}
\end{center}

\noindent
where $N$ is the number of nodal points.
\bigskip
\bigskip

Starting from the smooth three-folds on the right of the above figure we can
shrink N $S^3$'s to single points thereby ending up at a conifold point in the
moduli space. Since the Euler number of each point is unity we end up with a
singular three fold with an Euler number $\chi+ N$, where $\chi$ is the
Euler number of the initial 3-fold. Now we can replace each singular point with
a $S^2$ and obtain a new smooth 3-fold with an Euler number $\chi+ 2N$.
 This idea has been used by Strominger in 1995 in his
famous work on  conifold transition [7]. Strominger observed that the low
energy physics of
the moduli fields in the Calabi-Yau  compactifications of type II theories
is governed by some
$\sigma$ model targeted on the moduli spcae of the compactifying manifold.
As these fields vary in time they encounter a conifold point
thereby generating a singularity in the low energy theory. Strominger then
gave
a very nice physical understanding of these singularities. For the
sake of concreteness let us consider  type IIB compactification and
 a $D_3$ brane wrapped around
an $S^3$ in the compactifying Calabi-Yau three-fold. As the $S^3$ shrinks
to zero size the
3-fold approaches a conifold singularity. It also generates a state which
looks
like a black hole from the $4$-dimensional point of view.  Because of the
vanishing volume
of the wrapped brane this black hole will be massless and
will give rise to a massless multiplet in the low energy
effective 4-dimensional field theory. The origin of the singularity is the
appearance of
this massless multiplet.

\vskip 0.5cm
\noindent{\bf 4.  $D_3$ Branes and the AdS/CFT Correspondence}
\footnote{For a brief review see [12].  For  longer reviews see [13].}

A relationship like the one outlined in the previous section between gauge
theory and
gravity in $4$ dimensions would be highly interesting and will obviously
lead to
a deeper understanding of both gauge theory and gravitational physics. To
make  progress we
need to impose some
additional simplifying restrictions like supersymmetry, conformal symmetry
or rather super-conformal symmetry. With these restrictions some concrete
results have been obtained
by Gubser, Klebanov, Polyakov[15], Maldacena [2] and Witten
[15].
The starting point is the IIB string on Ad$S_5\times S_5$ background. These
authors have argued that
the ${ {\cal N}=4}$, { D=4}  super Yang-Mills on ${\rm I}\!{\rm R}^4$ is
equivalent
to IIB strings  on  $AdS_5\times S_5$.

>From the point of view of
our discussion in previous paragraphs we can say that for ${{\cal N}=4}$,
$D=4$ Yang-Mills theory the dual string theory  having the partition
function given in (1) is the type IIB superstring on the
$AdS_5\times S^5$ background. The background also needs to have a $RR$ charge
and a curvature related to $\lambda$. As we shall argue presently this duality
will be valid in the weak coupling limit of the string theory.

To give some more details of the gauge theory-gravity correspondence in
4-dimen\-sions we
need to go a little more
into the description of the $D$ brane physics. The semiclassical $D_3$
brane solution, which is a BPS state,
 is obtained by setting to zero
  the supersymetry variations of  fermionic fields in the type IIB
theory. The
$N$ parallel $D_3$ brane  configuration is given by,

$$ ds^2 = H^{- 1/2} (r) [ -dt^2 +dx^2 ] + H^{ 1/2} (r) [ dr^2 + r^2
g_{ij}dy^idy^j ]$$

where

$$ H(r) = 1+ {R^4 \over r^4},\qquad   R^4 = 4\pi g_sN \alpha^{\prime 2}$$

with $N$ an integer and $g_s$ the string coupling constant.
The part $ -dt^2 +dx^2$ defines a flat metric in  ${\Bbb R}^4$, while
$dr^2 + r^2 g_{ij}dy^idy^j  $ in general defines only a Ricci flat metric in a
$6$-dimensional transverse space  $Y_6$. The Ricci flatness of $Y_6$
requires that
the metric $g_{ij}dy^idy^j$ on the  $5$ manifold $X_5$  spanned by the
coordinates $y^i$ is
Einstein with a  cosmological constant equal to 4. Unless  $X_5 =S^5$ the
6-manifold  $Y_6$ is singular at $r=0$. In the singular case  $Y_6$ is called a
conifold because
 it admits a group of diffeomorphisms, $r\rightarrow tr$ with $t>0
$.  $Y_6$ is
a cone over $X_5$. The better known $D_3$ brane solution corresponds
to $Y_6= {\Bbb R}^6$ and $X_5 =S^5$.

 Note that as $r\rightarrow \infty$ the function $H$ approaches 1 and the
manifold
splits into ${\Bbb R}^4\times Y_6$. For  $Y_6 ={\Bbb R}^6$  we obtain the
10-dimensional
Minkowski
vacuum.  As $r\rightarrow 0$ we obtain $AdS_5\times X_5$. If
$Y_6 ={\Bbb R}^6$ we obtain the near horizon geometry $AdS_5\times S^5$.  We
thus
see that the brane configuration extrapolates between the vacuum Kaluza-Klein
type solutions at $r=0$ and $r=\infty$.

 The number of the unbroken supersymmetries depends on the choice of  $Y_6$.
For $Y_6 ={\Bbb R}^6$ only 16 of the original 32 real supersymmetries will
leave the
brane configuration invariant. For other choices of $Y_6$ the number of the
unbroken
supersymmetries will be smaller. One important feature, however, is that as
we approch
the limiting geometries at $r=0$ or $\infty$ both the bosonic and fermionic
symmetries
of the configuration increase. For arbitrary values of $r$ the bosonic
symmetries
are the 4-dimensional 10 parameter Poincar\'e group times a subgroup of
the isometry group of
$Y_6$ which leaves the function $H(r)$ invariant. For the case of $Y_6
={\Bbb R^6}$ this
is $SO(6)$. As we approach $r=0$ or $\infty$ the symmetry enlarges to
$SO(2,4)\times SO(6)$ ( near r =0) or the 55 parameter Poincar\'e
symmetries of ${\Bbb
R}^{10}$ (near $r=\infty$).
In this case the number of supersymmetries also increase to 32 in the two
extreme limits.

 The $r\rightarrow 0$ region always has the $AdS_5$ as part of its
geometry. Therefore,
 considering   $AdS_5\times X_5$ as a Kaluza-Klein type solution of  type IIB
supergravity
 the effective low energy theory will have the group $SO(2,4)$ as part of its
 symmetry group. $SO(2,4)$ is not only the isometry group of $AdS_5$ it is also
the group of conformal transformations of ${\Bbb R}^4$. The relevance of
${\Bbb R}^4$ to
this discussion
is due to the fact that it appears as the boundary of $AdS_5$. Along with
$SO(2,4)$,  the
isometry group of $X_5$ as well as the unbroken supersymmetries will be
contained
in the total symmetry group of the Kaluza Klein background $AdS_5\times X_5$.

 For the case of $X_5=S^5$ the low energy perturbative description of $N$
parallel $D_3$ branes is
in terms of ${\cal N}=4$, $U(N)$ supersymmetric Yang-Mills theory in ${\Bbb
R}^4$. In this case the total symmetry group is $SU(2,2|4)$.

 For the case of $Y_6={\Bbb R}^6$, the AdS/CFT correspondence asserts that,
within a
certain region of the parameter space, one can obtain nonperturbative
information
about the ${\cal N}=4$, $SU(N)$ Yang Mills theory from the low energy
type IIB supergravity on the $AdS_5\times S^5$ background. For this
description
to be a good approximation the radius $R$ of $S^5$ must be large with respect
to the $d=10$ Planck length. From the
above explicit formula for $R$ it is seen that a large value of $R^2$ (with
respect to $\alpha'$) requires
a large value for $g_sN$.  Since we are ignoring string loop effects we
should demand that $g_s$ is small. We will thus be dealing with a large $N$
gauge
theory in
4 dimensions.  A meaningful formulation of large $N$ gauge theories in
$d=4$ requires that the Yang-Mills coupling constant $g^2_{YM} =g_s$ must
decrease as we increase $N$ such that the 'tHooft coupling $\lambda = g_s
N$ remains
fixed. In our problem $\lambda$ has to be large. We  thus end up in the
regime of strong
'tHooft coupling limit of the $d=4$ super Yang-Mills theory, a domain in
which the
standard perturbative methods are not applicable.

This strong-weak duality also explains the discrepancy in the results of the
entropy calculations in Yang-Mills and string descriptions which were
puzzling when they
were first calculated. Here is a brief description of these results.
The black hole entropy
of a near extremal 3-brane of Hawking temperature $T$ turns out to be
[17]

$$
S_{BH}= {\pi^2\over 2} N^2 V_3 T^3
$$
where $N$ is interpreted as the charge of the brane and $V_3$ is the
spatial volume of the brane. On the other hand the entropy of a free gas
of particles in the $D=4$, ${\cal N} =4$  multiplet of
the $U(N)$ gauge theory turns out to be

$$
S= {2\pi^2\over 3} N^2 V_3 T^3 \ .
$$

 Note that the $V_3T^3$ behaviour is dictated by conformality of
the 4-dimensional  ${\cal N} =4$  super Yang Mills theory. The appearance
of $N^2$ is a reflection of degrees of freedom of $U(N)$,
as all the
fields
are in the adjoint representation of this group.  It is thus seen that
apart from a
relative factor of 3/4 the two entropies are the same. This discrepancy,
which created some confusion when it was first observed, is in fact
welcome, because,  we now know
 that the Yang Mills entropy  is obtained through a weak
coupling perturbative calculation. According to what we said above it is
the limit of the infinite 'tHooft coupling at 
which we expect to obtain agreements
with the supergravity calculations. It has been shown that in fact the
factor of 3/4 connects the two limits of the theory[18].

 At the level of  matching  the spectra of the Kaluza-Klein theory on
the maximally
symmetric $AdS_5\times S^5$
background with that of the ${\cal N}=4$ gauge theory on ${\Bbb R}^4$ the
 $AdS/CFT$ conjecture makes perfect sense. It essentially is a restatement
of the
symmetries of
the two theories [14].  The correspondence is less obvious when the background
has fewer
symmetries. It is  this less symmetric case which will be the focus of our
main interest in the following paragraphs.

In the maximally symmetric case every Kaluza-Klein mode on $S^5$ is dual to
an operator
in the Yang Mills theory. Let $J$ be one such mode and $\cal O$ its dual
operator
on the Yang-Mills side. A very explicit formulation of the correspondence
states that
the generating functional for the correlation functions of
 $\cal O$ can be obtained by evaluating the type IIB supergravity
partition function on the $AdS_5\times S^5$ background subject to the
boundary condition
that as we approach the boundary of $AdS_5$ the mode $J$ approaches a
boundary field
$\hat J$ [15]. The leading order contribution to the
generating functional of the connected Green's functions on the Yang-Mills side
will be given by the value of the classical
action of type IIB theory evaluated at a particular solution of equations
of motion
 subject to the given boundary conditions.

>From the brief sketch presented in the foregoing sections it becomes
apparent that the strong coupling limit of the Yang-Mills theory
in $D=4$ needs new degrees of freedom. Whereas the standard perturbative
analyses
are adequate for weak couplings,  the strong coupling domain of the same
gauge theory simplifies in the dual string description. From these analyses it
becomes also understandable why the attempts over the years to derive a stringy
behaviour from
the dynamics of Yang-Mills theory have failed. The reason obviously is that
the gravitational degrees of freedom were missing. Maldacena's conjecture
shows that it is the
throat region $(r<< R)$ of the $D_3$ brane geometry in $D=3$ dimensions
which should give a meaningful description of the strongly coupled gauge
theory in four space time dimensions.

\bigskip\bigskip

\noindent{\bf 5.  $D_3$ Branes Near Conifolds}
\bigskip

 One of the choices of $Y_6$ which leads to a less symmetric configuration is
the surface
$$Z_1Z_4-Z_2Z_3=0$$
 defined in ${\Bbb C}^4$. By a simple change of coordinates this surface
can also
be written as

$$\zeta_1^2 + ....+ \zeta_4^2 =0$$

which is the cone we encountered before. As we said earlier  its base $X_5$ is
obtained by intersecting the surface with the sphere

$$|\zeta_1|^2 + ....+ |\zeta_4|^2 =1$$

We can rotate the four variables $\zeta^i$ by elements of $O(4)$. This
group acts
transitively on the above intersection which defines the base of the cone.
The base $X_5$, therefore,  has to be a coset space. To
identify it consider some point, for example $ ( {1\over \sqrt 2}, 0,0,
{i\over \sqrt 2})$
and evaluate its stability subgroup, which is obviously the $O(2)$ subgroup of
$O(4)$ acting on the $2-3$ plane. We thus have $X_5= SO(4)/SO(2)$. $SO(4)$
invariant metrics on these manifolds have been worked in [6]. The metrics in
general
depend on two parameters $n$ and $n'$. It is customary to denote the
corresponding
manifolds by $T^{nn'}$. The explicit form of the metric on $T^{nn'}$ is,
 $${ds^4=c^2(dy_5-n\cos y_1dy_2-n'\cos y_3dy_4)^2+
a^2(dy_1^2+\sin^2y_1dy_2^2)+a^{{\prime}^2}(dy_3^2+\sin^2y_3dy_4^2),}$$

  All the coordinates $y^{\alpha}$ are angles, such that $ y=({y^1}, y^{2})$
parametrize
a $S^2$ of radius $a$ while $y' = (y^3 , y^4)$ parametrize a $S^2$ of
radius $a'$. The angle $y^5$ ranges from 0 to $4\pi$. The constant
$c$ is the
radius of the circle defined by $y^5$. The constants $n$ and $n'$
will be taken
to be integers.
Locally the manifold
 looks like $S^2\times S^2\times S^1$. Globally $T^{nn'}$ is a $U(1)$ bundle
over $S^2\times S^2$. The isometry
group is $SU(2)\times SU(2)\times U(1)$, where the $U(1)$ factor
is due to the translational invariance of the
coordinate $y^5$.

In order for $AdS_5\times T^{nn'}$ to be a supersymmetric solution of the type
IIB field equations it is necessary that
$a=a'= {1\over {\sqrt 6 \mid e\mid}}$, $n=\pm n'$ and $ec=-{1\over {3n}}$,
where, $e^2$ is related to the $AdS_5$ cosmological constant
through $R_{\mu\nu}= -4e^2g_{\mu\nu}$.
It has been argued
by Klebanov and Witten that the $U(1)$ factor should be identified with
 the $R$ symmetry of the world volume ${\cal N}=1$, $d=4$ superconformal
field theory
which arises as a consequence of the $AdS/CFT$ correspondence [5].  It has been
argued in [16] that
 there is a possible connection of the present theory with the $c=1$ non
critical string theory.
 In this context it becomes plausible that
the $R$ charge is
also  put in correspondence with the $U(1)$ group generated by the Liouville
mode of this theory. In this way the $R$ charges of the boundary
$d=4$ superconformal theory will be set in correspondence with the Liouville
momenta of the $c=1$ theory.

 We know from the perturbative description of a
collection of $N$ coinciding $D_3$ branes that the low energy description
 should be in terms of a super symmetric Yang-Mills theory on the
world volume of the brane.  What we are arriving at is that in the limit
of the strong 'tHooft coupling the type IIB superstring on
$AdS_5\times T^{1,1}$ is dual to the large $N$ limit of a super Yang-Mills
theory
on the world volume of the brane. In fact Klebanov and Witten [4] have
shown that the correct super Yang-Mills theory entering this duality is
 ${\cal N}=1$ superconformal $SU(N)\times SU(N)$ gauge theory on ${{\Bbb
R}}^{1,3}$, at a very particular
fixed
point which we are going to outline presently. Before doing this we note that
the background we are considering has a global symmetry of
$SU(2)_j\times SU(2)_\ell \times U(1)_R$. The Yang-Mills theory has scalar
chiral superfields $A_i$, $B_i$
and vector multiplets $W^1_\alpha$ and  $W^2_\alpha$ which belong to the
following representations of
$SU(N)\times SU(N)\times SU(2)_j \times SU(2)_l\times U(1)_R$

$$A_i\sim(N,\bar N; 2,1)_{1/2}\qquad B_i\sim(\bar N,N;1,2)_{1/2}$$
$$W^1_\alpha\sim(N^2-1,1;1,1)_1\qquad W^2_\alpha\sim (1,N^2-1;1,1)_1$$

Klebanov and Witten associate the following  conformal dimensions with
these operators

$$\Delta(A)=\Delta(B) ={3\over 4}$$
$$\Delta(W)={3\over 2}$$

Using the above super fields one can construct
 gauge invariant  chiral operators. Here are some examples,

$$\begin{array}{lll}{\rm Tr}(AB)^k  &  \Delta ={3\over 2}k  &   R=k\\[5mm]
{\rm Tr}(W_\alpha(AB)^k)  &  \Delta ={3\over 2} (k+1) &  R = k+1\\[5mm]
{\rm Tr}(W^\alpha W_\alpha (AB)^k)  &
\Delta =3+{3\over 2}k  & R=k+2\end{array}$$

For a more complete list see A. Ceresole et al. [19].
The question is how does one generate this spectrum of dimensions and $R$
charges on the supergravity side. The $AdS/CFT$ correspondence states that the
conformal dimensions of the chiral operators on the Yang-Mills side are given
in terms of the masses of the Kaluza-Klein modes.
For example for a  gauge invariant Yang-Mills operator dual to
a p-form the relation is as follows:
$$\Delta = 2+\sqrt{(4-p)^2+4(mR)^2}$$
 where $m=$ Kaluza-Klein mass of a mode originating from a $p$-form field
on $X_5$ and it has the
general form of ( at least for the case of $X_5=S^5)$
$m = {n\over R}$, with  $ n \in {\rm Z}\!\!{\rm Z}\,$.

 The dimensions  are thus independent of $R$ and therefore
they are also independent of the 'tHooft coupling.
 Likewise a gauge invariant Yang-Mills operator dual to a massive  spin 3/2
field on the supergravity side the formula is
$$\Delta = 2+\vert m+3/2\vert $$

It  becomes therefore important to find the masses of the Kaluza-Klein
modes.  This problem has been solved completely [16],[19].
To perform the spectral analysis one notices that
$T^{1,1}$ is a magnetic monopole bundle over $S^2\times S^2$.
 So if we expand  on the fibre coordinate $y_5$  we obtain fields defined
on $S^2\times S^2$, viz,

$$\Phi(y_{{\large 1}},y_{_{{\large  2}}};y_{_{{\large 3}}},
y_{_{{\large 4}}},y_{_{{\large 5}}})
=\mathop{\displaystyle\sum}\limits_{s\in{1\over2}\hbox{\Bbb Z}}
\Phi_{_{{\large  5}}} (y_{_{{\large 1}}},y_{_{{\large  2}}},y_{_{{\large 3}}},
y_{_{{\large   4}}}) e^{is\ y_{_{{\large  5}}}}$$

$\Phi_s$ are coupled to spin connections
on $S^2\times S^2$ as well as to magnetic monopole fields

$$A=\cos y_{_{{\large 1}}}\ dy_{_{{\large  2}}}\qquad A'=
\cos y_{_{{\large  3}}}\ dy_{_{{\large  4}}}$$

We can expand $\Phi_s$ on the basis of Wigner
functions on $S^2\times S^2$.
The spectrum is classified according to
 $SU(2)_\ell\times SU(2)_j\times U(1)_s$ representations
$(\ell,j;s)$. Following procedures developed in [20], one can obtain the
eigenvalues of various laplacian type operators on $T^{1,1}$. For example
the scalar Laplacian has  the following eigenvalues

$$H(s)=6\left(\ell(\ell+1)+j(j+1)-{s^2\over 2}\right)$$
$$\ell\geq \vert s\vert\qquad j\geq \vert s\vert$$

Likewise for the Dirac operator /$\!\!\!\! D$ we obtain

$${1\over 2}\pm \sqrt{H\left(s-{1\over 2}\right)+4}$$
$$-{1\over 2}\pm \sqrt{H\left(s+{1\over 2}\right)+4}$$

These eigenvalues give the masses of various AdS fields.
It must be noted that, since the eigenvalues of various Laplacians are
irrational functions of the quantum numbers,  the masses will also be
 irrational functions of $\ell$, $j$ and $s$.
Therefore the conformal dimensions  $\Delta$  which are equal to the
$ AdS$ energy become irrational functions of $\ell$, $j$ and $s$,
$$\Delta(\psi_\mu) =2+\vert m_{s/2}+3/2\vert$$
$$\Delta(\lambda) =2\pm\vert m_{1/2}\vert$$

We thus see that the boundary conformal field theory
contains infinite number of operators with irrational dimensions [21]. However,
most of the modes at the bottom of Kaluza-Klein towers have rational
dimensions.
For example
for the scalars such modes have the quantum numbers
$j=\ell=s$ and masses
$$m^2=H(s) =(3s+2)^2-4$$
 Their conformal dimensions will therefore be given by
$$\begin{array}{rcl}\Delta_\pm &=& 2\pm\sqrt{m^2+4}\\[3mm]
& = & 2 \pm (3s+2)\end{array}$$

These correspond to the chiral operators $Tr(AB)^k$ on the Yang-Mills side
$$\Delta = {3k\over 2}\quad \ell = j = {k\over 2}$$

For this particular example it is also instructive to work out
 the corresponding supergravity modes. To this end one writes the
metric  and the 4-form potential on $T^{(1,1)}$ as

$$g_{\alpha\beta} =\bar g_{\alpha\beta}+h_{\alpha\beta}$$
$$A_{\alpha_1\dots\alpha_4} =\bar
A_{\alpha_1\dots\alpha_4}+a_{\alpha_1\dots \alpha_4}$$
where $\bar g$ and $\bar A$ are background fields
The $ D = 10$ supergravity equations for these perturbations
leads to a $2\times 2$ sector involving
$h{^\alpha}_{\alpha}$  and $ a_{\alpha_1\dots \alpha_4}\ .$
Diagonalizing the linearized field equations one  obtains the masses

$$\begin{array}{rcl}m^2_\pm +4&=&\left(\sqrt{H+4}\pm 4\right)^2\\[3mm]
& = & \left( (3s+2)\pm 4\right)^2\end{array}$$
$$j = \ell = s$$

Substitute $m^2_-$ in $\Delta_+$
$$\begin{array}{rcl} \Delta_+ & = & 2+\sqrt{m^2_-+4}\\[3mm]
& = & 3s = 3{k\over 2}\quad {\rm if}\ s = {k\over 2}
\end{array}$$

Many other states with rational dimensions have been matched in both sides.
 They all fall into {\it short} multiplets of $SU(2,2\vert 1)$. Thus their
dimensions are protected.

\smallskip

 To summarize conifolds appear in many places in string theory like the $c=1$
non--critical strings, Chern Simons
theory on $S^3$, compactification of IIB and $D_3$--branes in IIB etc.
In the case of $D_3$--branes they lead to ${\cal N}=1$ superconformal
$SU(N)\times SU(N)$ Yang--Mills
on the boundary of AdS$_5$.
Many Kaluza-Klein modes have been matched with the {\it short} muliplets on
the YM side.
There are  infinite numbers of Kaluza Klein towers which lead to irrational
dimensions on the Yang--Mills side. This is unlike the AdS$_5\times S_5$ case.
The symmetry group is
$$SU(2,2\vert 1)\times SO(4)$$
One may wonder if the $c=1$ non--critical string theory plays a role in
the boundary Yang Mills theory?

\bigskip\bigskip

\noindent
{\bf 6. Relation to c=1 string theory}

\bigskip

The fundamental fields of the $c=1$ theory are two scalars $X$ and $\phi$
and the $b,c$ ghost fields.
$X$ is targeted on a $S^1$ while $\phi$ is coupled to a background charge.
At the self dual radius $R=1/\sqrt 2$ of the circle there is a $SU(2)\times
SU(2)$ symmetry.
The BRS cohomology classes are organized according to the representations
of this group.
 These classes are also labelled by their ghost numbers.
Of interest to us are the ghost number zero, one and two operators given
respectively
 by ${\cal O}_{s,p}(z) \overline{\cal O}_{s,p'}(\bar z)$,
$Y^+_{s+1,p}(z)\overline{\cal O}_{s,p'}(\bar z)$
 as well as $a(z,\bar z){\cal O}_{s,p}(z)\overline{\cal O}_{s,p'}(\bar z)$
and $Y^+_{s,p}(z) \bar Y^+_{s,p'}(z)$. Here
we follow the notation of [9].
The complex conjugates of these operators should also be added to the list.
In each case the subscript $s$ characterizes the $SU(2)\times SU(2)$
content of each object.
For a given integer or 1/2 integer $s$ the indices $p$ and $p'$ range from
$-s$ to $+s$.

Now consider a Kaluza-Klein tower originating from a $q$-form field in
$T^{11}$.
For a given $U(1)$ charge $s$ we consider the modes at the bottom of
each tower (those which presumably have rational conformal dimensions in
the boundary gauge theory).
An observation made in [16] is that these Kaluza-Klein modes are in
correspondence with
the ghost number $q$ cohomology classes in the $c=1$ theory.
Note that the modes originating from the components of the metric in $T^{11}$
(which are not $q$-forms in the internal space!) do not seem to have a
counterpart in the $c=1$ side.
Furthermore the modes corresponding to an operator containing
$a(z,\bar z)$ in the $c=1$ side can actually be gauged away in the
Kaluza-Klein side.

The $c=1$ theory has an infinite dimensional algebra given in terms of
the volume preserving diffeomorphisms of the quadric cone
$Z_1Z_2-Z_3Z_4=0$. As we mentioned in previous section
 the base of this cone is isomorphic to $T^{11}$.
In the context of present discussion the 3 complex dimensional Ricci flat
cone is in fact
identical to the subspace transverse to the $D_3$-brane solution of the
type IIB supergravity.
Our Kaluza-Klein background is a near horizon approximation to this $D_3$
brane geometry.
Thus the cone seems to be the common geometrical entity in the two very
different looking
theories\footnote{Similar remarks can be made about the manifolds $T^{nn}$.
In this
case we should consider the $c=1$ string theory at $n$ times the self-dual
radius.}.

The cone is singular at its apex and as we discussed in the previous section
one can resolve the singularity by deforming the defining
equation into $Z_1Z_2-Z_3Z_4=\mu$.  From the point of $c=1$ theory $\mu$
corresponds
to the 2-dimensional cosmological constant. One can also consider a
topological $\sigma$ model
targeted on a CY three fold near a conical singularity, for which the local
equation is
the same as our quadratic expression. For both of these theories the free
energies
can be evaluated as a function of $\mu$ and can be expressed as a genus
expansion.

In [8] Ghoshal and Vafa argued that in fact the two theories must be the same.
They observed that, at the self dual radius, the $g=0,1$ and 2
contributions to the
free energy of the $c=1$ theory agree with the corresponding terms of the free
energy of the topological sigma model near the conifold singularity.
Subsequently,
assuming the type II-heterotic duality, the results of [22] gave further
support to the Ghoshal-Vafa conjecture.
These authors calculated the coefficient of the term $R^2F^{2g-2}$
in the effective action of the heterotic theory compactified on
$K_3\times T^2$ and realized that for any $g$ the coefficient
is also given by the genus $g$ term of the partition function of the $c=1$
theory at the self dual radius. More recently Gopakumar and Vafa [23]
have calculated the $\sigma$-model partition function near a conifold
singularity
and have proven the conjecture made in [8].

A better understanding of the correspondences noted above
may require the unraveling of the relevance of the volume
preserving diffeomorphisms of the cone in the $D_3$ brane context.
On the basis of the observations made in this note we would like to
think that the $c=1$ theory at the self dual radius has a role
to play in organizing the chiral primaries of
the boundary $SU(N)\times SU(N)$ superconformal gauge theory.

\bigskip\bigskip
\noindent{\bf Acknowledgements}
\bigskip

I would like to thank Matthias Blau for useful discussions.
The very kind hospitality of the organizers of the Conference
on Quantum Symmetries in Goslar and  Wigner Symposium in Istanbul is
gratefully acknowledged.

\end{document}